\documentclass[pdftex,twocolumn,epjc3]{svjour3}          % twocolumn

\RequirePackage[T1]{fontenc}

\smartqed  % flush right qed marks, e.g. at end of proof

\RequirePackage{graphicx}
\RequirePackage{mathptmx}      % use Times fonts if available on your TeX system
\RequirePackage{flushend}
\RequirePackage[numbers,sort&compress]{natbib}
\RequirePackage[colorlinks,citecolor=blue,urlcolor=blue,linkcolor=blue]{hyperref}

\usepackage{multirow}
\usepackage{graphicx}
\usepackage{epstopdf}

\journalname{Eur. Phys. J. C}

\begin{document}

\title{Using wavelet analysis to compare the QCD prediction and experimental data on $R_{e^+e^-}$ and to determine parameters of the charmonium states above the $D\bar D$ threshold}

\author{V.K. Henner\thanksref{e1,addr1,addr2,addr3}
        \and
 C.L. Davis\thanksref{e2,addr1}
\and
 T.S. Belozerova\thanksref{addr2}
}
\thankstext{e1}{e-mail: vkhenn01@louisville.edu}
\thankstext{e2}{e-mail: c.l.davis@louisville.edu}

\institute{Department of Physics, University of Louisville, Louisville, KY 40292, USA\label{addr1}
          \and
          Department of Theoretical Physics, Perm State University, 614990 Perm, Russia\label{addr2}
          \and
Department of Mathematics, Perm State Technical University, 614990 Perm, Russia\label{addr3}
}

\date{Received: date / Accepted: date}

\maketitle

\begin{abstract}
The first part of our analysis uses the wavelet me\-thod to compare the Quantum Chromodynamic (QCD) 
prediction for the ratio of hadronic to muon cross sections in electron-positron collisions, $R$, 
with experimental data for $R$ over a center of mass energy range up to about 7 GeV. A direct comparison 
of the raw experimental data and the QCD prediction is difficult because the data have a wide range 
of structures and large statistical errors and the QCD description contains sharp quark-antiquark 
thresholds. However, a meaningful comparison can be made if a type of ``smearing'' procedure is used 
to smooth out rapid variations in both the theoretical and experimental values of $R$. 
A wavelet analysis (WA) can be used to achieve this smearing effect. The second part of the 
analysis concentrates on the 3.0 -- 6.0 GeV energy region which includes the relatively wide charmonium 
resonances $\psi(1^-)$. We use the wavelet methodology to distinguish these resonances from experimental 
noise, background and from each other, allowing a reliable determination of the parameters 
of these states. Both analyses are examples of the usefulness of WA in extracting information 
in a model independent way from high energy physics data. 
\end{abstract}

\section{Wavelet Transformations}

Let us start with a brief description of the continuous wave\-let transformation (WT). 
The WT of function $f(t)$  is defined by
\begin{eqnarray}
w(a,t)=\frac{1}{\sqrt{aC_{\varphi}}} \int\limits_{-\infty}^{+\infty}\varphi^{*}
\left(\frac{t'-t}{a}\right)f(t')dt',
\label{g1}
\end{eqnarray}
where $C_{\varphi}$ is a normalization constant subject to the choice of wavelet. The decomposition described 
by equation (\ref{g1}) is performed by convolution of the function $f(t)$ with a bi-parametric family 
of self-similar functions generated by dilatation and translation of the analyzing function $\varphi(t)$ 
called a wavelet,
\begin{eqnarray}
\varphi_{a,b}(t)=\varphi \left(\frac{t-b}{a}\right),
\label{g2}
\end{eqnarray}
where the scale parameter $a$ characterizes the dilatation, and $b$ characterizes the translation. 
It is a kind of ``window function'' with a non-constant window width. High frequency wavelets are narrow 
due to the factor $1/a$, while low frequency wavelets are much broader. The function $\varphi(t)$  
should be well localized in both time and Fourier space and must obey the admissibility condition,
$\int\limits_{-\infty}^{+\infty}\varphi(t)dt$. This condition requires $\varphi(t)$ must be an 
oscillatory function and, if the integral (\ref{g1}) converges, the completeness of the wavelet 
functions provides the existence of inverse transformation,
\begin{eqnarray}
f(t)=\frac{1}{\sqrt{C_{\varphi}}} \int\limits_{-\infty}^{+\infty}\int\limits_{0}^{+\infty}\varphi
\left(\frac{t-t'}{a}\right)w(a,t')\frac{dt'da}{a^{5/2}} dt.
\label{e3}
\end{eqnarray}

In contrast to Fourier analysis, the WT depends both on $t$ and the frequency providing 
an optimal compromise with the uncertainty principal. One of the advantages of wavelet analysis 
is a fairly low sensitivity of the restored signal to any physically reasonable continuation 
of the function $f(t)$  outside the interval $\left(t_{min},t_{max}\right)$  where the data 
are known. To fill in gaps between the experimental points we use a linear interpolation 
(different interpolations lead to minimal difference in the restored signal). Note that since 
the average value of any wavelet is zero, the mean value of the WT is zero, so that 
$\left< f\right>$  must be added to the reconstructed signal to restore the mean value of 
the original signal.

Wavelets with good localization and a small number of oscillations are commonly used to recognize 
the local features of data, and to find the parameters of dominating structures (location 
and scale/width). In this work, we use one of the most popular wavelets of this type, 
the so-called ``Mexican Hat'',
\begin{eqnarray}
\varphi(t)= \left( 1-t^2 \right ) e^{-t^2/2}.
\label{e4}
\end{eqnarray}
This wavelet is plotted in Figure 1 for three values of the scale parameter, $a = 1$, 2 and 0.5.

\begin{figure}[!h]
\centering
  \includegraphics[scale=0.65]{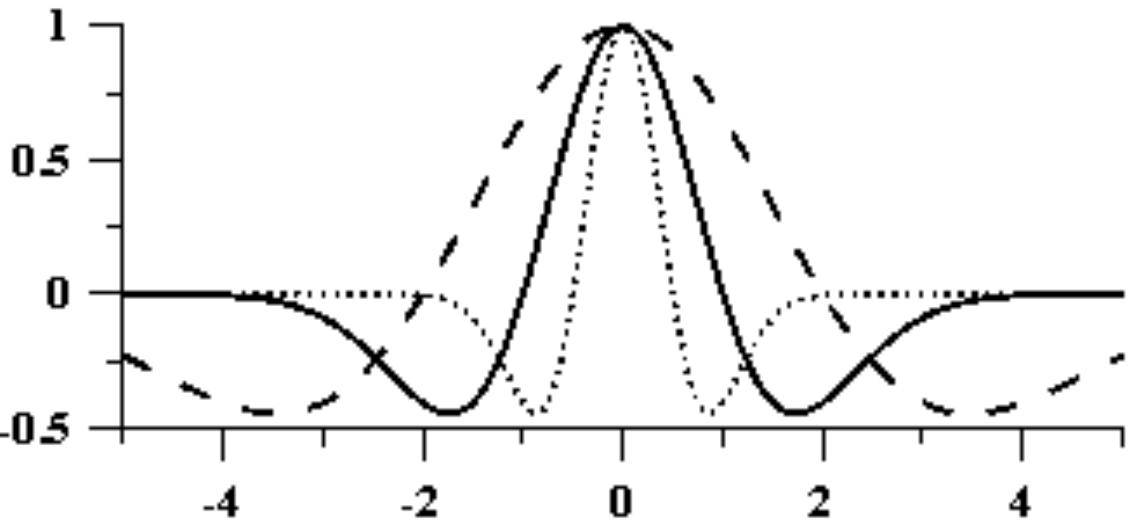}
\caption{Wavelet ``Mexican Hat'' $\varphi(t)= \left( 1-t^2 \right ) e^{-t^2/2}$. Solid line $-$ $\varphi(t)$, dashed line $-$ $\varphi(t/2)$, 
dotted line $-$ $\varphi(t/0.5)$.}
\label{fig:1}       % Give a unique label
\end{figure}

\section{Wavelets and the $R$ ratio}

Our first goal is to use wavelet methodology to compare the QCD prediction of $R$ to experimental 
data in the center of mass energy range up to about 7 GeV. In zeroth order in the strong coupling 
constant $\alpha_s$, the ratio $R$ is given by

\begin{eqnarray}
R(Q)=\frac{\sigma\left( e^+e^- \rightarrow hadrons \right )}
{\sigma\left( e^+e^- \rightarrow \mu^+\mu^- \right )} \approx 
3 \sum_{q} e^2_q \equiv R^{(0)}(Q),      
\label{e5}
\end{eqnarray}

\noindent
where the summation extends over the quark flavors $q=u,d,s,c$ available up to the center 
of mass energy $Q$. The QCD $\alpha_s$ corrections to $R^{(0)}(Q)$  can be presented 
in different forms. To be specific we adopt the form of reference [1],
\begin{eqnarray}
R(Q)= 3 \sum_{q} e^2_q T\left( \rm {v}_q \right ) \left [ 1+ g\left( \rm{v}_q \right ) {\cal R} \right ],      
\label{e6}
\end{eqnarray}                                          
where
\begin{eqnarray}
{\cal R} = \frac{\alpha_s}{\pi} 
\left [ 1+ \frac{\alpha_s}{\pi} + C_2 \left ( \frac{\alpha_s}{\pi} \right )^2 + C_3 \left ( \frac{\alpha_s}{\pi} \right )^3 + ... \right ],
\nonumber\\
{\rm v}_q = \left [1 - 4m_q^2 / Q^2 \right ] ^{1/2}, \; \;
T({\rm v}) = {\rm v} \left (3-{\rm v}^2 \right ) /2,
\nonumber\\
g({\rm v}) = \frac{4\pi}{3} \left [\frac{\pi}{2{\rm v}} - \frac{3 + {\rm v}}{4} \left ( \frac{\pi}{2} - \frac{3}{4\pi} \right )\right ]
\nonumber
\end{eqnarray}
\noindent
and the summation in (\ref{e6}) is over all quark flavors whose masses $m_q$ are less than 
$Q/2$. For this analysis we use the quark masses, the form of the coefficients $C_2$ and $C_3$ calculated in [3] 
and the energy dependence of $\alpha_s\left (Q^2 \right )$  from [2].

It is very hard to determine the role of these QCD corrections by comparison with 
experimental data for $R(Q)$ due to the large statistical errors and plethora of overlapping and 
interfering resonances in this region.  An additional complicating factor is that the QCD perturbative approach exhibits 
sharp quark-antiquark thresholds. However, a meaningful comparison can be made by applying some 
type of ``smearing'' procedure, which has the effect of smoothing out rapid variations in both 
the theoretical and experimental values of $R$.

Before describing our wavelet analysis of this data it is worth considering the methodology 
developed in references [1] and [2] to compare experimental data and QCD predictions.
The smearing procedure used in these analyses calculates a smeared ratio $R$ as follows,
\begin{eqnarray}
R(s,{\rm \Delta})= \frac{{\rm \Delta}}{\pi} \int \limits_{0}^{s_{max}} \frac{R(s')}
{\left(s'-s \right)^2 + {\rm \Delta}^2 } ds',      
\label{e7}
\end{eqnarray}                                       
\noindent
where $\sqrt{s}=Q$ is the square of the center of mass energy and ${\rm \Delta}$  
is the ``smearing'' parameter. In references [1] and [2] to evaluate the integral (\ref{e7}), 
it was necessary to exclude sharp resonances, such as the $\psi(3.100)$  and the much wider 
$\rho$  peak.  In addition, a term is added to account for the contribution from $s_{max}$ to 
$\infty$, assuming that $R$ remains constant above $s_{max}\approx 60 \; {\rm GeV}^2$. 
Originally, the smearing procedure in reference [1] supposes a global constant value 
${\rm \Delta} = 3 \; {\rm GeV}^2$  in (\ref{e7}). However it was found in reference [2] that 
for different energy regions it would be better to use different values of ${\rm \Delta}$. 
Note that the use of an energy dependent ${\rm \Delta}$ in (\ref{e7}) reflects the necessity 
for different treatment of different energy scales.

The WT methodology provides an alternative, model independent, smearing method, which does not 
require different treatment in differing energy scales. Under WT to separate the signal from 
the background noise, wavelet reconstruction is performed for scales greater than a 
certain scale $a_{noise}$  -- the boundary, or cut-off, scale [4]. In deciding on the 
appropriate boundary scale that will separate the noise-like high frequency components 
of the data we take a pragmatic line of reasoning. That is, the best choice for $a_{noise}$  
is the smallest value which will smooth out any rapid variations in the data enabling us 
to reproduce stable results for low frequencies (resonance area). A similar pragmatic 
strategy was applied in the analyses of [1] and [2] in choosing the parameter ${\rm \Delta}$  of equation (\ref{e7}). 
The best value of ${\rm \Delta}$ is large enough to compare the smeared $R$ with QCD models, but not so 
large that all the fine detail of the data is smoothed away.

\begin{figure}[!h]
\centering
  \includegraphics[scale=0.9]{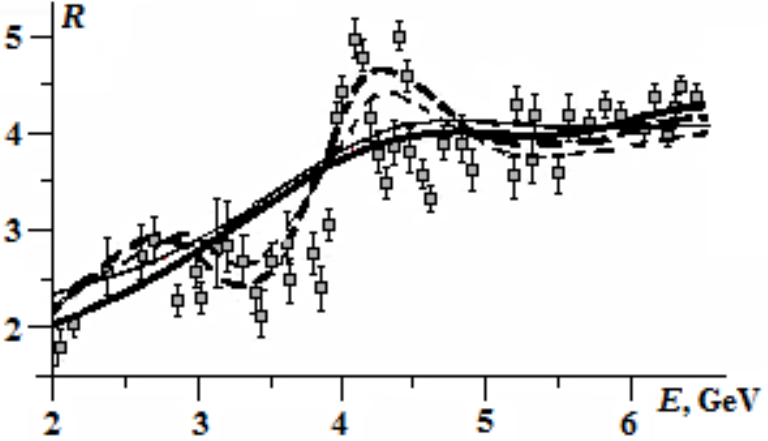}
\caption{Wavelet reconstruction of $R_{e^+e^-}$.  Dashed and solid bold curves correspond to wavelet reconstruction of the experimental data with $a_{noise} = 0.6$ and $a_{noise} = 1.2$, respectively.  Dashed and solid thin curves correspond to wavelet reconstruction of QCD calculations (\ref{e6}) with the same  $a_{noise}$ values.}

\label{fig:2}       % Give a unique label
\end{figure}

Figure 2 displays wavelet reconstructed (smeared) experimental data with cut-off values 
$a_{noise} = 0.6$ (bold dashed curve) and $a_{noise} = 1.2$ (bold solid curve). 
The data is a compilation of measurements from many different experiments obtained from the 
Particle Data Group (PDG) [5]. In contrast to the analyses of [1] and [2], under the WT approach 
there is no need to remove sharp resonances ($\psi(3.100)$  etc.) by hand, they 
effectively become part of the high frequency noise and the wavelet analysis smears them out. 
The thin dashed line and thin solid line are wavelet reconstructed QCD curves with 
$a_{noise} = 0.6$ and $a_{noise} = 1.2$, respectively. As can be seen, the two curves with $a_{noise} = 1.2$, representing 
experiment and theory, are in good agreement. It should be noted that the contribution 
to all of the restored data curves in Figure 2 from the data region above 6.5 GeV is negligible.

\section{ $\psi$ states above the $D\bar D$  threshold and the wavelet procedure}

Detailed information on the charmonium resonances $\psi(1^-)$ above the $D\bar D$ threshold 
at 3.73 GeV comes primarily from the measurement of $R_{e^+e^-}$. This data, provided by the 
Particle Physics Data Group (PDG) [5], comes from the work of many experimental collaborations 
over a period of more than 30 years. Several broad vector resonances are observed with varying 
degrees of clarity. The current best estimate of the masses and widths from the PDG are 
presented in Table 1.

\begin{table*}[!ht]
\caption{Properties of vector (charmonium) resonances above the charm threshold from the PDG and the BES [6] collaboration}
\begin{center}
\begin{tabular}{|c||c|c||c|c|}
\hline
\multirow{2}{*}{Resonance} & \multicolumn{2}{c||}{Mass (GeV)} & \multicolumn{2}{c|}{Full Width (GeV)}\\
\cline{2-5}
            & PDG [5] & BES [6] & PDG [5] & BES [6]\\
\hline
$\psi(3.770)$ & 3.773$\pm$ 0.0003 & 3.772 $\pm$ 0.002 & 0.027 $\pm$ 0.001 & 0.030 $\pm$ 0.009 \\\hline
$\psi(4.040)$ & 4.039 $\pm$ 0.001 & 4.040 $\pm$ 0.004 & 0.080 $\pm$ 0.010 & 0.085 $\pm$ 0.012 \\\hline
$\psi(4.160)$ & 4.191 $\pm$ 0.005 & 4.192 $\pm$ 0.007 & 0.103 $\pm$ 0.008 & 0.072 $\pm$ 0.012 \\\hline
$\psi(4.415)$ & 4.421 $\pm$ 0.004 & 4.415 $\pm$ 0.008 & 0.062 $\pm$ 0.020 & 0.072 $\pm$ 0.019 \\\hline
\end{tabular}
\end{center}
\end{table*}

The main difficulty encountered by each of the col\-la\-bo\-rations in making these measurements 
is large statistical errors and hence the difficulty of separating resonance ``signal'' 
from noise. There is significant disagreement between the collaborations on many of the 
resonance parameters.  Therefore, due to the possibility of systematic errors we do not 
combine measurements from different experiments, but  choose to base this initial analysis 
on data from the BES collaboration which exhibits clear evidence of four broad resonances 
above the $D\bar D$ threshold. In addition to the PDG values, Table 1 shows the masses 
and widths of these resonances from the most recent BES analysis [6].

Before analyzing the charmonium resonances parameters, we first determine the non-resonant 
background contribution using a WT followed by wavelet reconstruction.  This methodology, 
with its excellent scaling property, allowing the analysis of data with varying resolution, 
is ideally suited to separate  resonances from noise, background and each other.  A very helpful 
representation of the WT revealing all the features of the complete spectrum of the signal 
is the ``time- frequency'' plane (Figure 3(a)). This is a multi-resolution spectrogram, which shows the 
frequency (scale) contents of the signal as a function of energy. Each pixel on the spectrogram 
represents $w(a,t)$ for a particular $a$ (scale) and $t$ (in our case $t$  is energy, $E \equiv Q$).
The location of spots on the vertical axis (scale axis, $a$) corresponds to the width of the 
maximum. The intensity of dark spots shows the amplitudes of maxima. The WT image (wavelet plane) 
of the BES charmonium data obtained with the ``Mexican Hat'' wavelet is shown in panel (a) 
of Figure 3. The WT localizes the structures in a fashion that allows us to estimate the masses 
of the resonances and their widths -- all four $\psi$ resonances are clearly seen on the 
wavelet plane. The straight horizontal line corresponds to the boundary scale $a_{noise}$, 
which can be chosen to cut off small scale structures, which in this case corresponds to 
experimental noise.  By choosing a larger value of $a_{noise}$ it is possible to also cut 
out the resonances, leaving only the background. The background curves for three substantially 
different choices of $a_{noise}$ (0.15, 0.25 and 0.35) are displayed in Figure 3(b). In the 
following analysis we show that the charmonium resonance parameters are not sensitive 
to the choice of $a_{noise}$.

\begin{figure}[!h]
\centering
  \includegraphics[scale=0.65]{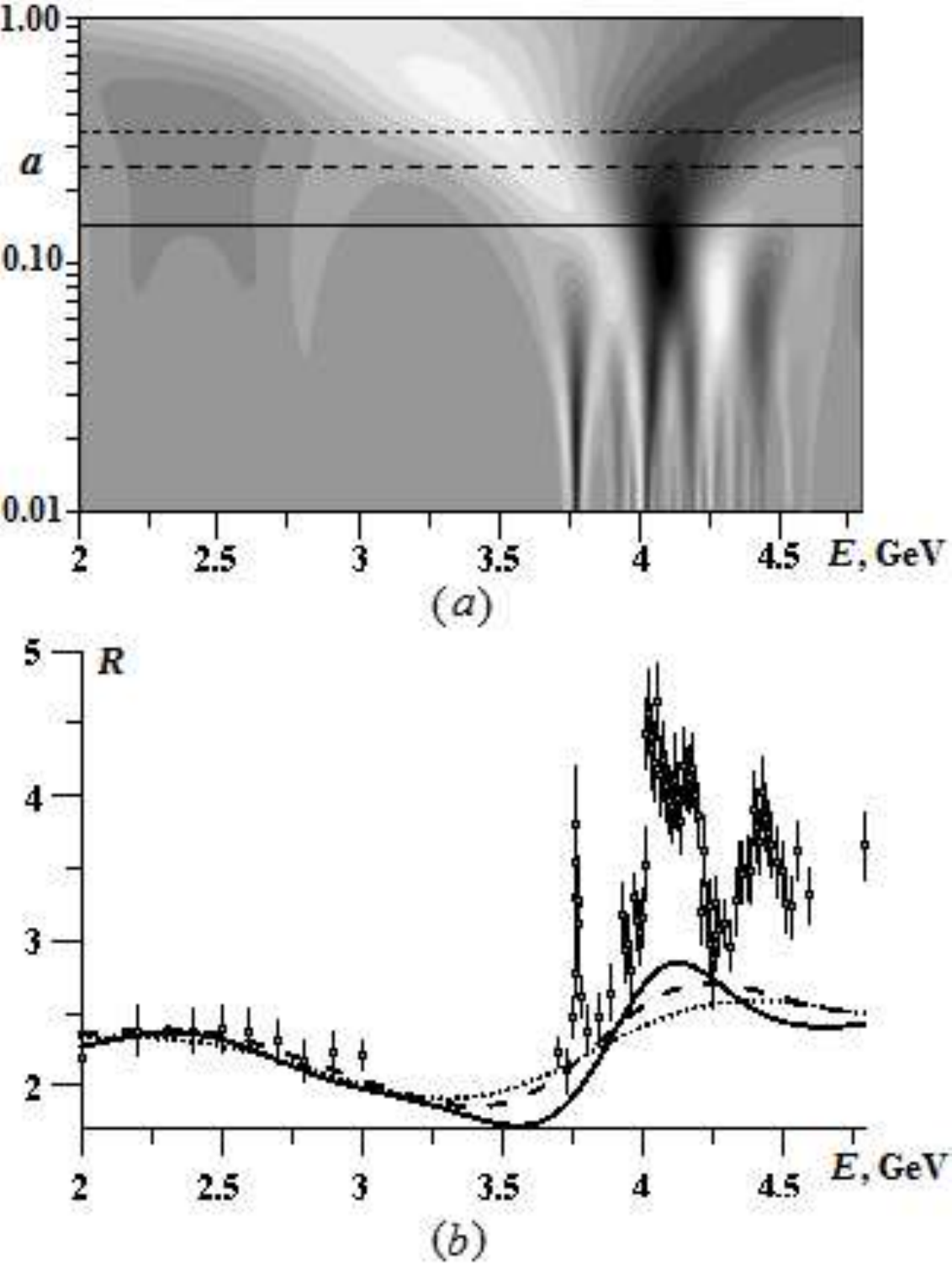}
\caption{ (a) Wavelet plane for $a \in [0.01,1]$; (b) background for $a_{noise} = 0.15$ (solid line),
$a_{noise} = 0.25$ (dashed line), $a_{noise} = 0.35$ (dotted line).}
\label{fig:3}       % Give a unique label
\end{figure}

For each value of $a_{noise}$ we subtract this ``wavelet background'' from the raw experimental 
data leaving only the contribution from the broad charmonium resonances. We then perform 
a least squares fit of the ``signal'' in this energy range to the sum of four Breit-Wigner 
resonances of the form,
\begin{eqnarray}
R_{res} = \frac{9}{\alpha_{em}^{2}} \sum \limits_{r=1}^4 B_{lr} B_{hr} \frac{M_{r}^{2} \Gamma_{r}^{2}}
{\left ( s-M_{r}^{2} \right )^2 + M_{r}^{2} \Gamma_{r}^{2}},      
\label{e8}
\end{eqnarray}
\noindent 
where $M_{r}$, $\Gamma_{r}$, $B_{lr}$ and $B_{hr}$  are, respectively, the mass, total width, 
leptonic branching fraction and hadronic branching fraction of the resonance.  Each Breit-Wigner 
has three fitted parameters, $M$, $\Gamma$ and the product of $B_{lr}$ and $B_{hr}$. It is worth 
noting that since these resonances partially overlap, if their decay channels are specified, 
we can improve their resolution by using the multi-channel unitary scheme described 
in reference [7].  We defer this investigation to a future analysis.

The values of the fitted parameters for each of the four Breit-Wigner resonances for 
backgrounds obtained with
three different values of $a_{noise}$ are presented in Tables 2-4.  
The chi squared per degree of freedom for the three fits are 1.11, 1.22 and 0.99 
for the $a_{noise}$ values of 0.15, 0.25 and 0.35, respectively.

\begin{table}[!h]
\centering
\caption{Breit-Wigner fitted parameters for $a_{noise} = 0.15$}
\label{tab2}
%\begin{tabular}{\columnwidth}{@{\extracolsep{\fill}}cccc@{}}
\begin{tabular}{cccc}
\hline\noalign{\smallskip}
Resonance & Mass (GeV) & Width (GeV) & $B_l B_h \cdot 10^5$\\
\noalign{\smallskip}\hline\noalign{\smallskip}
$\psi(3.770)$  & 3.772 $\pm$ 0.0002  & 0.032 $\pm$ 0.005  & 0.983 $\pm$ 0.076 \\
$\psi(4.040)$  & 4.042 $\pm$ 0.001 & 0.088 $\pm$ 0.013 & 0.921 $\pm$ 0.067 \\
$\psi(4.160)$  & 4.161 $\pm$ 0.006 & 0.100 $\pm$ 0.019 & 0.664 $\pm$ 0.069 \\
$\psi(4.415)$  & 4.430 $\pm$ 0.007 & 0.098 $\pm$ 0.014 & 0.936 $\pm$ 0.071 \\
\noalign{\smallskip}\hline
\end{tabular}
\end{table}

\begin{table}[!h]
\centering
\caption{Breit-Wigner fitted parameters for $a_{noise} = 0.25$}
\label{tab3}
%\begin{tabular}{\columnwidth}{@{\extracolsep{\fill}}cccc@{}}
\begin{tabular}{cccc}
\hline\noalign{\smallskip}
Resonance & Mass (GeV) & Width (GeV) & $B_l B_h \cdot 10^5$\\
\noalign{\smallskip}\hline\noalign{\smallskip}
$\psi(3.770)$  & 3.772 $\pm$ 0.0001 & 0.022 $\pm$ 0.004  & 0.888 $\pm$ 0.094 \\
$\psi(4.040)$  & 4.043 $\pm$ 0.003 & 0.097 $\pm$ 0.011 & 1.110 $\pm$ 0.064 \\
$\psi(4.160)$  & 4.185 $\pm$ 0.005 & 0.083 $\pm$ 0.016 & 0.715 $\pm$ 0.077 \\
$\psi(4.415)$  & 4.423 $\pm$ 0.004 & 0.090 $\pm$ 0.016 & 0.819 $\pm$ 0.076 \\
\noalign{\smallskip}\hline
\end{tabular}
\end{table}

\begin{table}[!h]
\centering
\caption{Breit-Wigner fitted parameters for $a_{noise} = 0.35$}
\label{tab4}
%\begin{tabular}{\columnwidth}{@{\extracolsep{\fill}}cccc@{}}
\begin{tabular}{cccc}
\hline\noalign{\smallskip}
Resonance & Mass (GeV) & Width (GeV) & $B_l B_h \cdot 10^5$\\
\noalign{\smallskip}\hline\noalign{\smallskip}
$\psi(3.770)$  & 3.773 $\pm$ 0.0001 & 0.022 $\pm$ 0.004 & 0.888 $\pm$ 0.090 \\
$\psi(4.040)$  & 4.043  $\pm$ 0.003 & 0.097 $\pm$ 0.011 & 1.155 $\pm$ 0.061 \\
$\psi(4.160)$  & 4.165  $\pm$ 0.004 & 0.083 $\pm$ 0.013 & 0.907 $\pm$ 0.073 \\
$\psi(4.415)$  & 4.423  $\pm$ 0.004 & 0.090 $\pm$ 0.014 & 0.889 $\pm$ 0.071 \\
\noalign{\smallskip}\hline
\end{tabular}
\end{table}

As can be seen in Tables 2-4, the fitted parameters of the four charmonium resonances 
are not significantly different for the three values of $a_{noise}$.  Therefore, in Figure 4 
we present only the fitted curve (\ref{e8}) for $a_{noise}= 0.35$. 

It should be noted that in this analysis, after 
subtracting the wavelet background, we exclude the four highest energy data points from the 
fitting procedure.  Without these points the $\psi(4.415)$  resonance is ``well-shaped''; 
if these four points are included the resonance is no longer ``well-shaped'' and its width 
is about twice that of the values presented in Table 1.

\begin{figure}[!h]
\centering
  \includegraphics[scale=0.65]{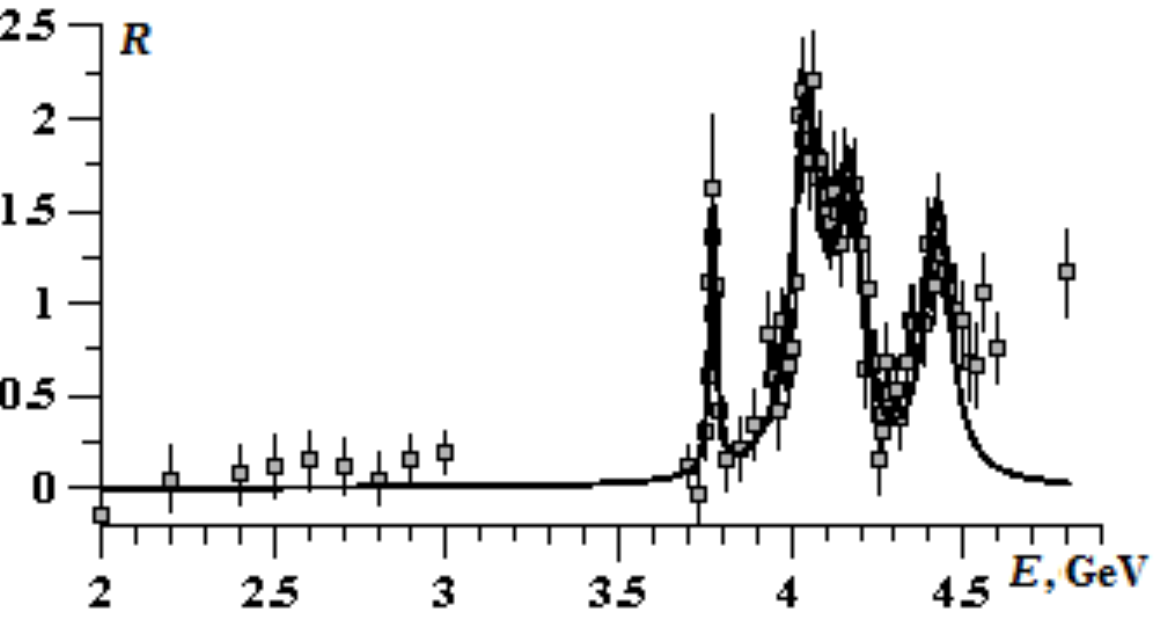}
\caption{Breit-Wigner fit of BES data with ``wavelet background'' subtracted, $a_{noise}= 0.35$}
\label{fig:4}       % Give a unique label
\end{figure}

Justification for excluding the four highest energy points can be made as follows.
If instead of the wavelet fitted background we fit the data with four Breit-Wigners and a 
parabolic (or linear) background, including the four highest energy data points, we see that the 
fourth resonance is not ``well-shaped'' and the last few data points appear to be associated with 
the background rather than the 4th resonance, see Figure 5.  Furthermore, the fitted width of this 
fourth resonance (Table 5) is significantly larger than that of the second and third resonances and 
the published PDG value.  In order to show the association of the highest energy points with the 
parabolic background we do not subtract the background in Figure 5.  It should be emphasized 
that inclusion or exclusion of these four data points does not significantly alter the background 
curve obtained via the wavelet method and that the choice of a polynomial background is arbitrary, 
whereas the wavelet background is obtained from the data itself in a model independent manner.

\begin{table}[!h]
\centering
\caption{Breit-Wigner parameters for fit with parabolic background}
\label{tab5}
%\begin{tabular}{\columnwidth}{@{\extracolsep{\fill}}cccc@{}}
\begin{tabular}{cccc}
\hline
Resonance & Mass (GeV) & Width (GeV) & $B_l B_h \cdot 10^5$\\
\hline
$\psi(3.770)$  & 3.770  $\pm$ 0.001 &  0.021 $\pm$  0.004 & 1.202  $\pm$ 0.097  \\
$\psi(4.040)$  & 4.048  $\pm$ 0.003 &  0.107 $\pm$  0.010 & 1.946  $\pm$ 0.079  \\
$\psi(4.160)$  & 4.166  $\pm$ 0.003 &  0.086 $\pm$  0.013 & 1.335  $\pm$ 0.067  \\
$\psi(4.415)$  & 4.426  $\pm$ 0.005 &  0.148 $\pm$  0.022 & 1.001  $\pm$ 0.085 \\
\noalign{\smallskip}\hline
\end{tabular}
\end{table}

\begin{figure}[!h]
\centering
  \includegraphics[scale=0.65]{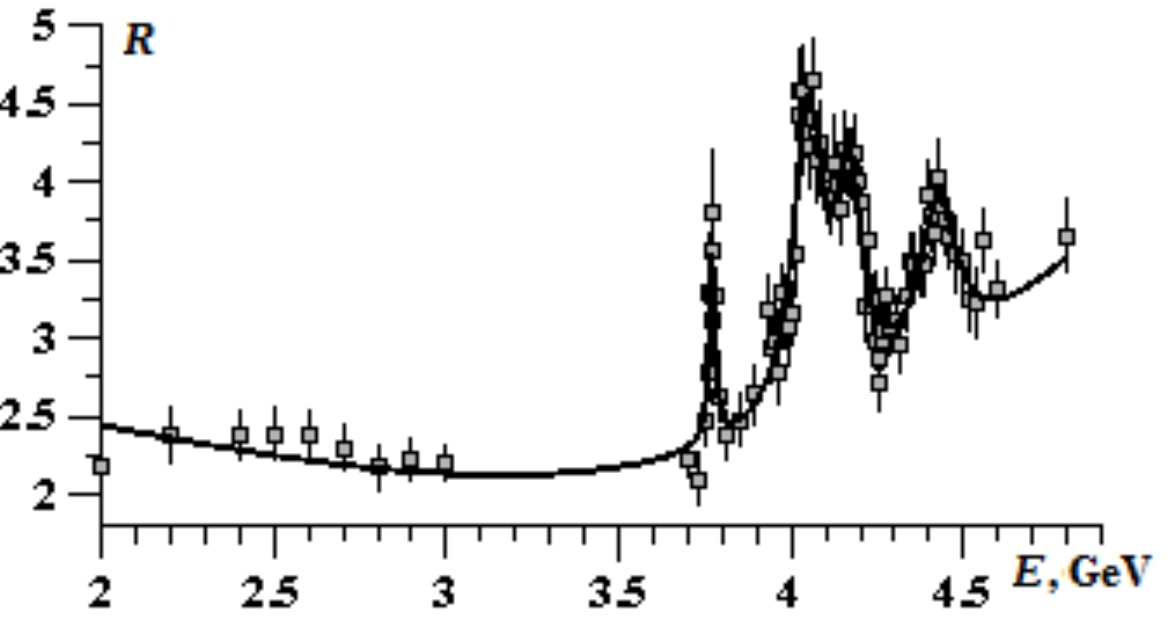}
\caption{Breit-Wigner and parabolic background fit of BES data (contrary to Fig.4 the background is not 
subtracted from the data).}
\label{fig:5}       % Give a unique label
\end{figure}

To demonstrate the reliability of our WT methodology in extracting the parameters of 
these charmonium resonances we performed a study using simulated data as follows. 
Starting from the PDG values of masses and widths of the four charmonium resonances 
given in Table 1 we generate a spectrum of four Breit-Wigners of the basic form of equation (\ref{e8}).  
To simulate real experimental data we discretize the spectrum from 2.0 to 5.5 GeV in 0.05 GeV 
intervals, then add a random vertical shift from -0.2 to +0.2 and random error bars in the range 
0.15 to 0.4 to the data points.  To complete the simulated data spectrum a representative 
background curve is added to the data points.  Finally we pass this simulated data through 
the same WT analysis chain applied to the real data.  This gives us a wavelet background 
curve and the masses and widths of the Breit-Wigner resonances after the wavelet background 
subtracted fit.  In addition to applying our WT methodology to the simulated data we perform 
a Breit-Wigner and parabolic background fit in exactly the same way as we did for the real data.

This study was performed for several different initial background curves.  In all such cases 
the WT methodology accurately reproduced the input background shape.  Fitting the wavelet 
background subtracted simulated data, as described above, also accurately reproduces the 
masses and widths of the assigned Breit-Wigner resonances, independent of the particular 
input background curve.  The parabolic background least squares fit of the simulated data 
was also able to reproduce the Breit-Wigner masses and widths, but in contrast led to fitted 
backgrounds very different from the input background.  We believe the fact that the masses 
and widths were accurately reproduced, despite the fitted background being very different 
from the input background, is largely due to the clean nature of the simulated data.  Real 
experimental data is clearly more complex, in which case the inability of the standard parabolic 
background least squares fit to extract an accurate background could lead to less reliable 
resonance parameters.

\section{Conclusion}

Experimental measurements of the ratio, $R$, comprise a wide range of structures with large 
statistical errors, making direct comparison with the predictions of QCD very difficult. 
However, a meaningful comparison can be made provided that some kind of ``smearing'' procedure, 
similar to that described in [1], is used to smooth out rapid variations in $R$. A wavelet 
analysis can be used to achieve this smearing effect. We compare the WT of the predictions 
of perturbative QCD and experimental $R$ data. The wavelet reconstruction of the $R$ experimental 
data preserves its main features, but with damped statistical errors and threshold singularities. 
The WT of QCD perturbation theory is in good general agreement with the WT experimental data.

Using the wavelet methodology to obtain the back\-gro\-und in the charmonium energy range above 
the naked charm threshold provides an important, model independent, alternative to other 
accepted methods. The masses and widths of the four vector mesons above the charm threshold,
$\psi(3.700)$,  $\psi(4.040)$, $\psi(4.160)$  and  $\psi(4.415)$, obtained from fitting wa\-ve\-let 
background subtracted data from the BES experiment, are found to be largely insensitive 
to the specific choice of the WT parameters and consistent with the BES and PDG reported 
values.\\

\begin{acknowledgements}
The authors wish to thank Harrison Simrall and Adam Redwine for their work on the fitting algorithm and A. L.Kataev for useful advice.
\end{acknowledgements}

\end{document}